\documentclass[10pt,aps,prd,twocolumn,
nofootinbib,noshowkeys,noshowpacs,
superscriptaddress,floatfix]{revtex4-2}
\usepackage{amsmath,amsfonts,amsthm,amssymb,tensor}
\usepackage[dvips]{graphics,graphicx}
\usepackage[usenames,dvipsnames]{color}
\definecolor{darkblue}{RGB}{0,0,196}
\definecolor{darkgreen}{RGB}{0,120,0}
\usepackage[colorlinks=true,linktocpage=true,linkcolor=darkblue,citecolor=red,urlcolor=darkblue]{hyperref}
\usepackage{cancel}
\usepackage{bbold}
\usepackage{subfigure}
\usepackage{stix}
\usepackage{multirow}
\usepackage{longtable}
\usepackage{color}
\usepackage[normalem]{ulem}
\usepackage{hyperref}
\usepackage{bigints}
\usepackage{xparse}
\usepackage{physics}
\usepackage{verbatim}
\usepackage{minibox}
\usepackage{comment}
\usepackage{appendix}
\usepackage{scalerel}
\usepackage{marginnote}
\usepackage{graphicx}
\usepackage[nice]{nicefrac}
\usepackage{soul}

\usepackage{hepunits}
\begin{document}
\preprint{}

\title{Study of angular observables in exclusive semileptonic \texorpdfstring{$B_c$}{} decays}
\author{Sonali Patnaik}
\email{sonali\_patnaik@niser.ac.in}
\affiliation{National Institute of Science Education and Research, An OCC of Homi Bhabha National Institute, Bhubaneswar, Odisha, India}
\affiliation{Theoretical Physics Department, CERN, 1211 Geneva 23, Switzerland}

\author{Lopamudra Nayak}
\email{lopalmn95@gmail.com}
\affiliation{National Institute of Science Education and Research, An OCC of Homi Bhabha National Institute, Bhubaneswar, Odisha, India}

\author{Priyanka Sadangi}
\email{priyanka.sadangi@niser.sc.in}
\affiliation{National Institute of Science Education and Research, An OCC of Homi Bhabha National Institute, Bhubaneswar, Odisha, India}

\author{Sanjay Swain}
\email{sanjay@niser.ac.in}
\affiliation{National Institute of Science Education and Research, An OCC of Homi Bhabha National Institute, Bhubaneswar, Odisha, India}
\author{Rajeev Singh}
\email{rajeevofficial24@gmail.com}
\affiliation{National Institute of Science Education and Research, An OCC of Homi Bhabha National Institute, Bhubaneswar, Odisha, India}
\affiliation{Center for Nuclear Theory, Department of Physics and Astronomy, Stony Brook University, Stony Brook, New York, 11794-3800, USA}
%
\begin{abstract}
In this work, we investigate angular observables such as the longitudinal polarization of charged leptons, $\tau$-polarization, and forward-backward asymmetry in semileptonic $B_c$ decays. Additionally, we provide predictions for lepton flavor violating observables, the $\mathcal{R}$ ratios in the decay channels $B_c \rightarrow \eta_c (J/\psi) l \nu_l$ and $B_c \rightarrow D (D^*) l \nu_l$ across the entire $q^2$ region. Our analysis is conducted within the Relativistic Independent Quark Model, focusing on the potential model-dependent aspects of these observables. We compare our model predictions with existing lattice predictions, highlighting the strong applicability of our framework in describing $B_c$ decays. Considering the forthcoming experimental upgrades and the Run 3 data results on $B_c$ meson decays, rapid confirmation of these quantities could indicate significant discoveries of physics beyond the Standard Model. This may open up new avenues for understanding the complex flavor dynamics in heavy meson decays.
    \end{abstract}
    \date{\today}




\maketitle
\section{Introduction}
\label{sec:intro}
The Standard Model (SM) treats charged leptons ($e^-$, $\mu^-$, $\tau^-$) as universal across the three generations, aside from kinematic effects due to their differing masses. Predictions of electroweak interactions within the SM exhibit Lepton Flavor Universality (LFU)~\cite{HFLAV:2016hnz,Bifani:2018zmi,Gambino:2020jvv,Bernlochner:2021vlv}, which has been experimentally verified in muon decays~\cite{BESIII:2013csc,NA62:2012lny,PiENu:2015seu}, tau decays~\cite{ParticleDataGroup:2022pth}, and Z boson decays~\cite{ALEPH:2005ab}. LFU is considered an accidental symmetry in the SM, emerging from the Yukawa interaction between leptons and the Higgs field, resulting in different lepton masses ($m_{\tau} > m_{\mu} > m_e$). This symmetry implies that physical processes involving charged leptons should exhibit LFU in observables such as decay rates and scattering cross-sections. However, LFU is violated (LFUV) by the charged lepton mass terms, which is observed in flavor-changing charged current (FCCC) transitions, mediated solely by charged weak bosons $W^\pm$.

The anomalies in charged current transitions, specifically involving $b \rightarrow c$ transitions, provide valuable opportunities to search for new physics (NP) beyond the SM. A persistent discrepancy exists in the world averages of the LFU ratios, which are defined as
\begin{equation}
\mathcal{R}_{D^{(*)}}=\frac{\mathcal{B} \left(B \rightarrow D^{(*)} \tau {\bar\nu_{\tau}}\right)}{\mathcal{B} \left(B \rightarrow D^{(*)} l {\bar\nu_l}\right)}\,, \qquad \text{with}\qquad \ell= \mu,e\,,
\end{equation}
where $D^{(*)}$ refers to $D$ and $(D^*)$ meson and $\mathcal{B}$ means the branching ratio. In March 2024, the LHCb collaboration announced preliminary results for $\mathcal{R}_D$ and $\mathcal{R}_{D^*}$ using the semileptonic-tagging method on a partial Run 2 dataset \cite{LHCb24:2024LHCb}. Recently, the observable $\mathcal{R}_{D^{(*)}}$ has deviated more than three standard deviations ($\sigma$) from the SM predictions which has been reported in \cite{Chen:2024hln}. Additionally, the Belle II collaboration reported its first preliminary results for $\mathcal{R}_D$ and $\mathcal{R}_{D^*}$ using the semileptonic tagging method ($\tau \rightarrow l \bar {\nu} \nu$)~\cite{BelleII24:2024BelleII}. An excess of over 3$\sigma$ has been observed in lepton universality tests in $B$ decays. Although the experimental uncertainty in the Belle II results is still large, it is expected to decrease significantly in the near future \cite{Belle-II:2022cgf}. This deviation has recently sparked significant interest within the particle physics community, prompting investigations into whether this tension is indicative of NP~\cite{ILCInternationalDevelopmentTeam:2022izu,Chen:2024hln}.

Similarly, evidence of LFU breaking has emerged in heavy-flavored mesons, particularly in $B_c$ meson decays, which is the primary focus of our current study. The $B_c$ meson, composed of a charm and a bottom quark, represents the lowest bound state of two heavy quarks with distinct flavors. Structurally, the $B_c$ meson shares similarities with charmonium ($c \bar c$) and bottomonium ($b \bar b$) states in terms of its bound state properties. Unlike double heavy quarkonia with hidden flavors, which can decay strongly and electromagnetically, the $B_c$ meson primarily decays weakly due to its position below the $B \bar D$ threshold. This characteristic makes it an ideal system for exploring heavy quark dynamics. Consequently, the $B_c$ meson exhibits a long lifetime. 
Given that both constituent quarks ($b, c$) are heavy, they decay through various weak channels with significant branching fractions. Despite being intermediate in mass and size between charmonium and bottomonium families, where interactions involving heavy quarks are well understood, many aspects of weak interactions within the $B_c$ sector remain obscured due to limited experimental data.

Among the various decay modes of the $B_c$ meson, the semileptonic decay mode holds particular significance. Its discovery at Fermilab by the CDF Collaboration \cite{CDF:1998axz} involved the analysis of the decay $B_c \rightarrow J/\psi l \nu_l$, where $J/\psi$ decays into a muon pair. Beyond extracting precise values of the Cabibbo-Kobayashi-Maskawa (CKM) matrix elements, studying semileptonic decays allows for investigating the universality of the coupling of the three charged leptons in electroweak interactions. This serves as a robust tool for testing the SM and probing potential effects of physics beyond the SM (BSM). Due to their straightforward theoretical description through tree-level processes in the SM, analyzing semileptonic decays helps disentangle the influences of strong interactions from those of weak interactions into a set of Lorentz invariant form factors. Therefore, investigating semileptonic decay processes primarily involves computing the relevant form factors within a suitable phenomenological theoretical approach.

The source of tension with the SM predictions involving the semileptonic $B_c \rightarrow J/\psi$ decay is in the ratio
$\mathcal{R}(J/\psi)$, which is defined as
\begin{equation}
\mathcal{R}_{J/\psi}=\frac{\mathcal{B} \left(B_c{^+} \rightarrow J/\psi \tau^+ \nu_{\tau}\right)}{\mathcal{B}\left(B_c{^+} \rightarrow J/\psi \mu^+ \nu_{\mu}\right)}\,, 
\end{equation}
with $\tau^+$ decaying leptonically to $\mu^+$ $\nu_{\mu}$ ${\bar \nu_{\tau}}$.
The LHCb experiment measured $\mathcal{R}_{J/{\psi}} = 0.71 \pm 0.17 \pm 0.18$ \cite{LHCb:2017vlu}. For the $\mathcal{R}_{J/\psi}$ analysis, data from proton-proton ($pp$) collisions corresponding to 3 fb$^{-1}$ were collected at center-of-mass energies $\sqrt{s} = 7$ TeV and $8$ TeV \cite{LHCb:2017vlu}. The analysis involved reconstructing the $\tau$ lepton and performing a global fit on the missing mass squared ($q^2$) and decay time of $B_c$. Across all cases, the decay branching fractions consistently deviate from SM predictions, currently lying in the range of 0.25--0.28 \cite{Ivanov:2006ib,Hernandez:2006gt,Watanabe:2017mip}. The observed values are approximately 2$\sigma$ higher than the SM predictions, as shown in Table~\ref{table:2}. The variation in SM results arises from different modeling approaches for the form factors \cite{Hernandez:2006gt,Ivanov:2006ni}. The excess of $\mathcal{R}_{J/\psi}$ over SM predictions not only offers further insights into the unresolved $\mathcal{R}_{D(*)}$ puzzle but also suggests the need to consider potential interactions involving new particles in $B_c \rightarrow J/\psi l \nu_l$.

Recently, the CMS Collaboration also released the initial preliminary outcome of $\mathcal{R}_{J/\psi}$ using the B parking data~\cite{CMS24:2024CMS} 
\begin{equation}
    \mathcal{R}_{J/\psi}^{\text{CMS}} = 0.17^{+ 0.18}_{- 0.17\; \text{Stat.}} \,^{+ 0.21}_{\;- 0.22\; \text{Syst.}} \,^{+ 0.19}_{\;- 0.18\; \text{Theory}}\,.
\end{equation}
By naively averaging, the CMS experiment obtains,
 \begin{align}
     \mathcal{R}_{J/\psi} = 0.52 \pm 0.20 \,.
 \end{align}
This finding aligns with the SM prediction \cite{Harrison:2020nrv} within a margin of 1.3$\sigma$. Despite the significant experimental uncertainties in these results, it would be beneficial in the future to examine various scenarios to address the anomalies observed in $\mathcal{R}_{D^{(*)}}$. 

The recent Belle measurements of the longitudinal $\tau$ polarization, $P_{\tau}(D^*)$ \cite{Belle:2016dyj, Belle:2019ewo}, and the fraction of $D^*$ longitudinal polarization, $F_L^{D^*}$, in $B \rightarrow D^* \tau \nu_{\tau}$ decays have also garnered considerable attention in the investigation of these crucial observables
\begin{eqnarray}
P_{\tau}(D^*) &=& - 0.38 \pm 0.51 \,({\rm stat.})^{0.21}_{-0.16}\, ({\rm syst.})\,,\nonumber\\
F_{L}(D^*) &=& 0.60 \pm 0.08 \,({\rm stat.}) \pm (0.04)\,({\rm syst.})\,,
\end{eqnarray}
whereas the SM predictions of these observables are~\cite{Tanaka:2012nw,Huang:2018nnq,Bhattacharya:2018kig}
\begin{eqnarray}
P_{\tau}(D^*) &=& - 0.497 \pm 0.013\,,\nonumber\\
F_L(D^*) &=& 0.441 \pm 0.006 \quad {\rm or} \quad 0.457 \pm 0.010\,.
\end{eqnarray}
The experimental results seem to be consistent with SM predictions. Recently, the LHCb experiment \cite{LHCb:2023ssl} measured the fraction of longitudinal polarization of the $D^*$ meson in decays such as $B^0 \rightarrow D^{-} \tau^+ \nu_\tau$, where the $\tau$ lepton decays into three charged pions and a neutrino. The average value across the entire $q^2$ range is
\begin{eqnarray}
F_L^{D^*} = 0.43 \pm 0.06 \pm 0.03\,.
\end{eqnarray}
These results are in agreement with both SM predictions and findings from the Belle experiment. In $b \rightarrow c \tau \nu$ decays, the subsequent decay of the $\tau$ within the detector allows for the measurement of the $\tau$ polarization fraction.  This polarization fraction depends on the hadronic final state and is sensitive to contributions from BSM physics, offering a complementary probe alongside branching ratios or differential distributions of the three-body final state.

To address these established scenarios and assess the status of BSM physics, various methodologies \cite{Tanaka:2012nw, Watanabe:2017mip, Bhattacharya:2018kig, Korner:1989ve} are under consideration. These observables play a critical role in constraining and determining plausible extensions of the SM. Particularly noteworthy are the $\tau$ polarization, longitudinal polarization $\left(P_L\right)$, forward-backward asymmetry parameters $\left(\mathcal{A}_{\rm FB}\right)$, along with kinematic distributions, polarization asymmetries, and other LFUV parameters, collectively forming a comprehensive set for rigorous testing of the SM.

This motivates us to present predictions for these observables for $B_c \rightarrow \eta_c (J/\psi) l \nu_l$ and $B_c \rightarrow D (D^*) l \nu_l$ transitions. We refine the numerical formulations for $\mathcal{R}_{J/\psi}$, $\mathcal{R}_{\eta_c}$, $\mathcal{R}_D$, and $\mathcal{R}_{D^*}$ within sizable uncertainties. In addition to the $\mathcal{R}$ ratios, we introduce predictions for angular observables for the first time, including $\tau$ polarization $\left(P_{\tau}(\eta_c), P_{\tau}(J/\psi), P_{\tau}(D), P_{\tau}(D^*)\right)$ of the final state leptons, and the Forward-Backward asymmetry $\left(\mathcal{A}_{\rm FB}\right)$ across the entire physical range of momentum transfer squared, $q^2$, a feature absent in our previous study \cite{Nayak:2021djn}. Similar to the initial measurement of $P_{\tau}^{D^*}$ at Belle \cite{Belle:2016dyj, Belle:2019ewo}, the upcoming LHCb experiment could potentially measure $P_{\tau}(\eta_c)$, $P_{\tau}(J/\psi)$, $P_{\tau}(D)$, and $P_{\tau}(D^*)$ in $B_c$ meson decays.

Theoretically, the form factors for the semileptonic $B_c \rightarrow J/\psi$ transitions across the entire $q^2$ region have recently been computed using Lattice QCD (LQCD) simulations \cite{Harrison:2020gvo,Harrison:2020nrv}. Theoretical predictions for $B_c \rightarrow \eta_c l \nu_l$ decays are also available through two approaches: Perturbative QCD (PQCD) and PQCD combined with LQCD~\cite{Hu:2019qcn}. More recently, LQCD has calculated the branching fraction ratio for $B_c \rightarrow D l \nu_l$ \cite{Cooper:2021bkt}. Accurate SM predictions for the normalization and shape of the form factors in the semileptonic $B_c \to D (D^*)$ decays will improve experimental observations of this process and potentially provide a new exclusive determination of the CKM matrix element $|V_{ub}|$.
The LHCb collaboration anticipates that with Upgrade II \cite{LHCb:2020pro}, it will achieve sufficient precision to measure $B_c \to D^{0} l \nu_l$ competitively, offering a precise determination of $V_{ub}$. 

Our understanding of $B_c$ form factors significantly lags behind that of $\bar B^0 \rightarrow D^*$ transitions, primarily due to the scarcity of experimental data for $B_c$ decays and the involvement of two heavy quark flavors in both the initial $(b \bar c)$ and final $(c \bar c)$ states. This combination disrupts heavy quark symmetry (HQS), although residual heavy quark spin symmetry (HQSS) remains applicable, reducing the number of form factors under the infinite heavy quark limit \cite{Branz:2009cd,Kiselev:1999sc}. Unlike in the case of $\bar B^0 \rightarrow D^*$ transitions, HQSS does not determine the normalization of the form factors as HQS does. Theoretical extraction of form factors directly from first principles of QCD has proven challenging due to its non-abelian and non-perturbative characteristics. As an alternative, phenomenological models are employed to describe the bound-state nature of hadrons and their decay properties. Transition form factors that parameterize semileptonic decay amplitudes are computed using integrals involving overlapping meson wave functions derived from various theoretical approaches. These methodologies encompass the potential model approach \cite{Chang:1992pt}, Bethe-Salpeter approach \cite{Liu:1997hr,AbdEl-Hady:1999jux}, QCD sum rules, non-relativistic QCD (NRQCD) \cite{Kiselev:1999sc,Kiselev:2000pp}, relativistic quark model based on quasi-potential approach \cite{Ebert:2003cn}, non-relativistic quark model approach \cite{Hernandez:2006gt}, Bauer-Stech-Wirbel framework \cite{Dhir:2008hh}, PQCD approach \cite{Wang:2012lrc,Rui:2016opu}, covariant confined quark model approach \cite{Ivanov:2000aj,Issadykov:2017wlb,Tran:2018kuv}, and LQCD approach \cite{Harrison:2020gvo,Harrison:2020nrv}.

A quark potential model is successful when it accurately reproduces observed data across various hadron sectors. Despite the Lorentz structure used in the interaction potential, a phenomenological model is deemed reliable if it captures constituent-level dynamics within the hadron core and predicts a range of hadronic properties, including decay characteristics. However, the parameterization process in potential models involves some degree of arbitrariness. Thus, the potential model approach is not unique, particularly when constrained to reproducing experimental data within a limited range. Hence, extending the applicability of quark models to a broader spectrum of observed data is crucial.

In our analysis, we therefore adopt a potential model framework known as the Relativistic Independent Quark (RIQ) model, extensively utilized to describe a wide array of hadronic phenomena~\footnote{This model has been applied to study static properties of hadrons \cite{Barik:1987zb,Barik:1993aw}, as well as their decay properties, including radiative, weak radiative, rare radiative decays \cite{Barik:1994vd,Priyadarsini:2016tiu}; leptonic, weak leptonic, radiative leptonic decays \cite{Barik:1993yj,Barik:1993aw}; and non-leptonic decays \cite{Barik:2001vp,Kar:2013fna,Nayak:2022qaq}. Moreover, the RIQ model has been successfully applied to analyze semileptonic decays of heavily flavored mesons \cite{Barik:1996xf,Barik:2009zz}.}.
The decay modes $B_c\to \eta_c(J/\psi)l\nu_l$ are induced by the quark-level transition $b\to c l \nu_l$. Similarly, $B_c\to D(D^*)l\nu_l$ modes are induced by $b\to u l \nu_l$ at the quark level. The kinematic range of momentum transfer squared for the former class of modes is $0\leq q^2 \leq 10\ {\rm GeV}^2$, whereas for the latter type of decay modes, it extends to $0\leq q^2 \leq 18\ {\rm GeV}^2$. In the rest frame of the parent $B_c$ meson, the maximum recoil momenta of the final state charmonium ($\eta_c, J/\psi$) or charm mesons ($D, D^*$) can be estimated to be of the same order of magnitude as their masses. Given these kinematic constraints, analyzing the decay modes $B_c\to \eta_c(J/\psi)l\nu_l$ and $B_c\to D(D^*)l\nu_l$ is particularly intriguing. Some key points of our model are:
\begin{itemize}
    \item In our RIQ model, we calculate the relevant form factors across the entire kinematic range of  $q^2$, which enhances the accuracy of our predictions. In contrast, some of the aforementioned theoretical approaches determine form factors initially with endpoint normalization at either the minimum $q^2$ (maximum recoil point) or maximum $q^2$ (minimum recoil point). These form factors are then extrapolated phenomenologically to the entire physical region using monopoles, dipoles, or Gaussian ansatz, which can compromise the reliability of form factor estimates.
    \item We assess the relevant hadronic current form factors (both vector and scalar components) to analyze all potential semileptonic $B_c$-meson decay modes induced by $b\to cl\nu_l$ and $b \to u l \nu_l$ transitions at the quark level. This approach allows us to investigate the effects of lepton masses in semileptonic $B_c$-decays and predict observables in comparison with other SM predictions.
\end{itemize}
Our study focuses on recent findings in $B_c$-hadron decays, examining the applicability of the RIQ model framework in explaining LFU ratios and crucial angular observables, and comparing it with standard theoretical approaches. Despite inconsistencies in overall normalization, qualitative agreements are observed for observables based on distribution ratios such as the forward-backward parameter and $\tau$-polarization. Ratios between predictions in $\tau$ and $(e, \mu)$ modes, are also studied. Given the absence of lattice predictions for angular observables in $D$ and $D^*$ modes, our model's findings could provide valuable insights for upcoming experimental observations in $B_c$ decays.

This paper is structured as follows:
Section~\ref{sec:exp_outlook} provides an overview of current experimental prospects regarding LFUV and angular observables.
Section~\ref{sec:physical_obs} defines the physical observables analyzed in this study.
Sections~\ref{sec:RIQM} and \ref{sec:testsofLFUV} delve into the RIQ model framework and the general kinematics for investigating semileptonic decays $b \rightarrow c,u$.
Section~\ref{sec:num_res} presents the numerical analysis and various findings concerning $B_c$ decays.
Finally, section~\ref{sec:conclusions} offers a brief summary and outlines future directions.
\section{Experimental outlook on LFUV and angular  observables}
\label{sec:exp_outlook}
LFU tests at LHCb reveal disparities in the hardware trigger efficiencies for electrons and muons. Despite the low production rate of $B_c$ mesons, LHCb continues to study various $B$-hadron species. Currently, systematic uncertainties in LFU and angular observable measurements are smaller than statistical uncertainties, which are expected to decrease further with additional data from Run 3. The ongoing commissioning of the LHCb Upgrade I detector aims to increase instantaneous luminosity by a factor of 5, targeting the collection of approximately 50 fb$^{-1}$ during Run 3. The proposed LHCb Upgrade II
aims to accumulate 300 fb$^{-1}$ throughout the HL-LHC era, enabling a detailed exploration of flavor-physics observables with unprecedented precision \cite{LHCb:2022ine}.

Recent Belle II measurements \cite{Forti:2022mti}, including a 40\% improvement in the statistical precision of $\mathcal{R}_{D^*}$ compared to Belle, show better alignment with SM predictions. With data from LHC Run 3, Belle II aims to provide the $\mathcal{R}_D$ measurement with uncertainties 2 to 3 times smaller than the current world average. Apart from these LFU measurement discrepancies, polarization observables also give us significant and non-trivial information. This is due to the possibility that these observables will enable us to identify the specific NP structure causing these deviations. The first observation by Belle Collaboration \cite{Belle:2016dyj,Belle:2019ewo} of $\tau$-longitudinal polarization of $D^*$, and fraction of the $D^*$ longitudinal mode in $\bar B \to D^{(*)} \tau \nu$, has open up a new window in the search for NP in semitauonic $B$ decays, which has prompted us to predict similar observables in other decay channels. Belle II's exceptional capability to reconstruct final states with missing energy and efficiently identify all types of leptons that will significantly enhance the understanding of these anomalies \cite{Belle-II:2022cgf}. 

Additionally, since 2019, the CMS Collaboration has implemented an innovative data recording method known as ``B Parking'' \cite{Bainbridge:2020pgi,CMS:Bpark}. While their first official results for $\mathcal{R}_{D^*}$ are still pending, it is expected that the experimental uncertainty will be comparable to that of other B factories. The CMS collaboration \cite{CMS:2024syx} has also updated the numerical formula for $\mathcal{R}_{J/\psi}$ to include general NP contributions and provided a prediction based on their fit study. Despite the large experimental uncertainties in these results, they will be valuable for testing certain NP scenarios in relation to the $\mathcal{R}_{D^*}$ anomalies in the future.

Experiments at ATLAS \cite{ATLAS:2008xda} actively explore BSM physics through both direct and indirect methods. Indirect searches focus on rare and forbidden decays of heavy mesons and baryons, where NP contributions can be significant. While the potential to measure these anomalies at ATLAS is yet to be demonstrated, making future contributions uncertain, the large datasets anticipated during LHC Run 3 may provide initial insights into these exceptionally exotic and challenging processes. This could help shed light on the perplexing anomalies and BSM physics.
\section{Physical Observables}
\label{sec:physical_obs}
We now list the set of observables which we will use in our subsequent study and phenomenological
discussion in the semileptonic $b \rightarrow c (u)$ transitions.
\begin{itemize}
    \item \emph{ Ratios of branching fraction (LFU observable)}: The first observable is the ratios of total branching fraction ($\mathcal{B}_{\rm tot}$) which is most commonly considered in experimental searches. The formula is given by
    \begin{equation}
    \mathcal{B}_{\rm tot} = \int_{m_l^2}^{(M-m)^2} \left(\frac{d\mathcal{B}\left(q^2\right)}{dq^2}\right)\;dq^2\,,
    \end{equation}
  which can be computed using \eqref{eq:dw}. Here, $m_l$ denotes the mass of the charged lepton ($e, \mu, \tau$), $M$ represents the mass of the parent meson ($B_c$) and $m$ represents the mass of final state hadrons. Using the above expression the LFU observable is defined as:
\begin{equation}
   \mathcal{R}_X = \frac{\mathcal{B} (B_c \to X \tau \nu_{\tau})}{\mathcal{B} (B_c \to X l \nu_{l})}, \quad \text{where $l = e, \mu$}\,.
\end{equation}
 $X$ is final state hadron
    \item \emph{Lepton-polarization asymmetry}: A study of the decay to a charged lepton with a specific polarization state allows for the measurement of lepton-polarization asymmetry, which is crucial for understanding the unique physics impacts of new particles. The longitudinal polarization observable $P_L$ is defined as:
    \vspace{-0.5mm}
    \begin{equation}
    P_L = \frac{\frac{d{\Gamma}_i}{dq^2}^{h=+1/2} - \frac{d\tilde{\Gamma}_i}{dq^2}^{h=-1/2}}{\frac{d{\Gamma}_i}{dq^2}^{h=+1/2} + \frac{d\tilde{\Gamma}_i}{dq^2}^{h=-1/2}}\,.
    \end{equation}

Here, the rates ${d\Gamma_i}/{dq^2}$ and ${d \tilde{\Gamma}_i}/{dq^2}$ correspond to the neutrino-lepton spin no-flip (in the limit of vanishing lepton mass) and flip (in the case of non-vanishing lepton mass) contributions, respectively. The combination of flip and no-flip contributions determines the longitudinal polarization $P_L$ of the leptons. Here, $h$ represents the helicity of the charged leptons, $h = s \cdot p = \pm 1/2$, where the sign corresponds to the spin polarization being parallel (+) or anti-parallel (-) to the lepton's momentum direction. For a massless lepton, the direction of its momentum relative to its spin is the same in every reference frame, resulting in positive helicity. However, for massive particles, a faster moving reference frame reverses the helicity.

    The polarization asymmetry is typically described using a standard set of helicity structure functions ($U, L, S$), which represent unpolarized-transversed, longitudinal, and scalar functions, respectively. The un-tilde quantities are for no-flip contributions, the tilde quantities are for flip contributions. 
    See section \ref{sec:testsofLFUV} for more details.
    \begin{equation}
    P_L = \frac{U + L - \tilde{U} - \tilde{L}-\tilde{S}}{U + L + \tilde{U} + \tilde{L} + \tilde{S}}\,.
    \end{equation}
     \item \emph{Forward-Backward asymmetry}: Another important quantity that could provide insights into the interactions of new particles is the forward-backward asymmetry of the lepton $l$, which is defined as:
    \begin{eqnarray}
  \frac{d\mathcal{A}_{\rm FB}(q^2)}{dq^2} &=& \frac{1}{{\mathcal{B}}_{\rm tot}} 
  \Bigg[\int_{0}^{1} d\cos\theta_l \frac{d\mathcal{B}}{dq^2 \,d \cos\theta_l} \,,\nonumber\\
  &&- \int_{-1}^{0} d\cos\theta_l \frac{d\mathcal{B}}{dq^2 d \cos\theta_l}\Bigg],
    \end{eqnarray}
    where $\mathcal{B} = \mathcal{B^+} + \mathcal{B^-}$. The angle $\theta_l$ is the polar angle between the emitted lepton and the $W$ boson in the $l \nu$ frame. This observable is normalized to the total branching fraction. The expression refers to the $q^2$-dependent quantity, with its integrated characteristic obtained after integration over the full $q^2$ range. We introduce $\mathcal{A}_{\rm FB}$ in terms of helicity structure functions: unpolarized $(U)$, longitudinal $(L)$, parity-odd $(P)$, scalar $(S)$, and scalar-longitudinal interference $(SL)$ as:
\begin{equation}
\mathcal{A}_{\rm FB} = \frac{3}{4}\left[\frac{\pm P + 4 \tilde{SL}}{U + \tilde{U} + L + \tilde{L} + \tilde{S}}\right]\,.
\end{equation}
\end{itemize}

\begin{figure}
\begin{center}
\includegraphics[width=0.45\textwidth]{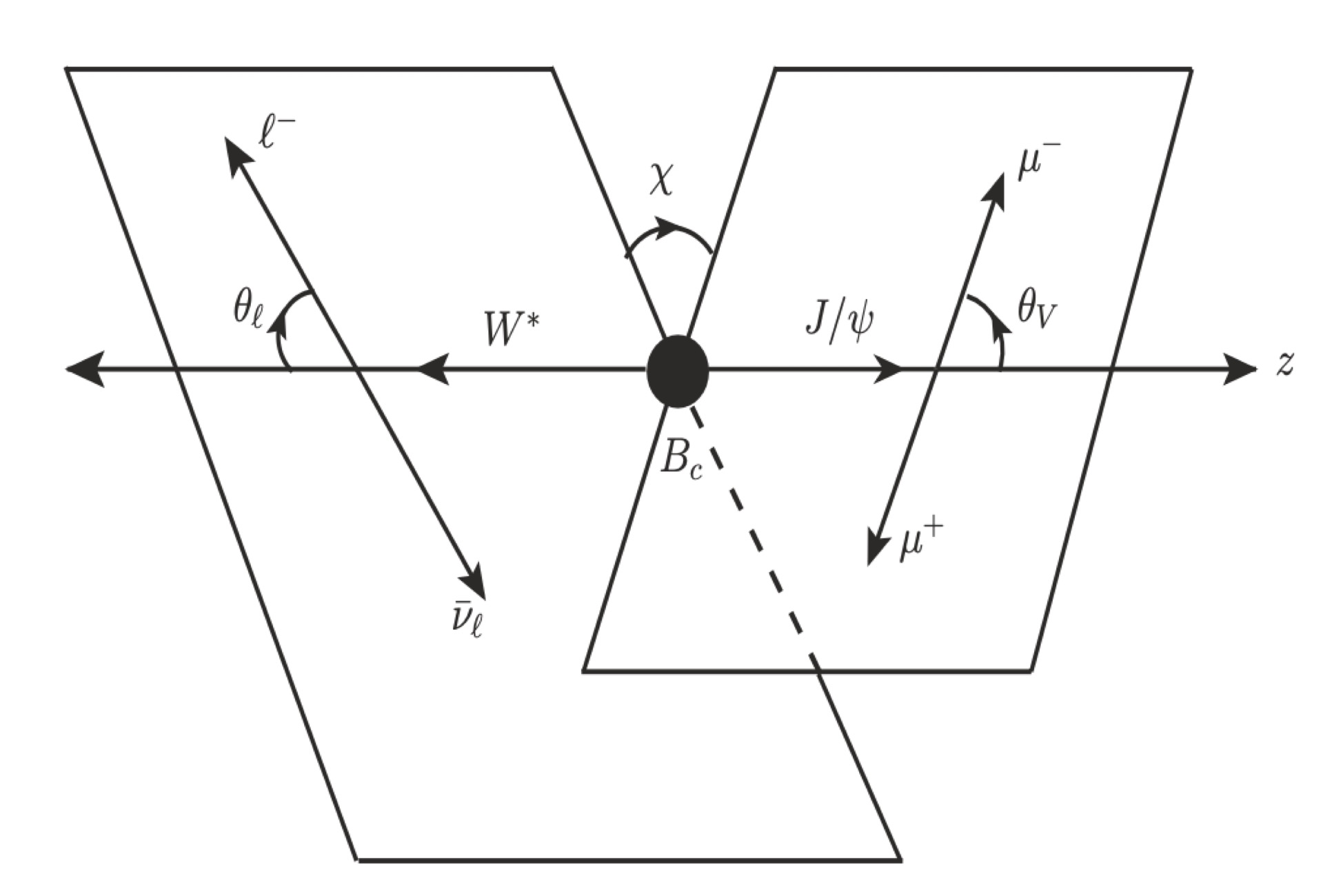}
\end{center}
\caption{Angular convention for $B_c \rightarrow X l \nu_l$ where $X = \eta_c , J/\psi$ , $D(D^*)$.}	
\label{fig:ad}
 \end{figure}
\section{RIQ Model}
\label{sec:RIQM}
\begin{figure}
\begin{center}
\includegraphics[width=0.45\textwidth]{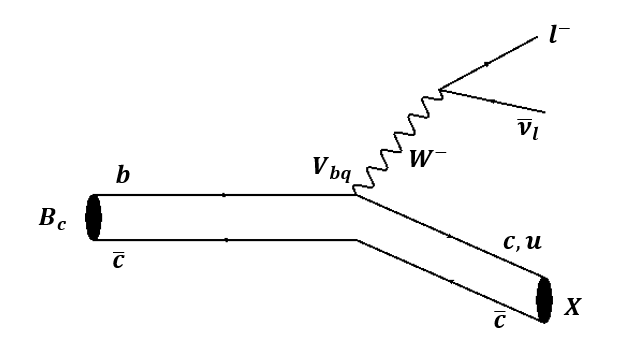}
\end{center}
\caption{SM contribution for $B_c \rightarrow X l \nu_l$ where $X = \eta_c(J/\psi)$ and, $D(D^*)$.}	
\label{fig:1}
 \end{figure}
Studying exclusive semileptonic decays involving non-perturbative hadronic matrix elements is a significant challenge. Currently, the rigorous application of field-theoretic techniques and QCD from first principles for reliable measurements of invariant transition amplitudes is not feasible. As a result, various theoretical approaches resort to phenomenological models to explore non-perturbative QCD dynamics. These methodologies are crucial for shaping differential decay amplitudes and facilitating sensitive measurements at the cutting edge of physics.


In this context, we present a concise overview of world averages using a model-dependent approach, specifically adopting the Relativistic Independent Quark Model (RIQM). The RIQM is based on a confining harmonic potential in the equally mixed scalar-vector form~\cite{Patnaik:2019jho,Patnaik:2022moy,Patnaik:2023ins}
\begin{equation}
U(r)=\frac{1}{2}\left(1+\gamma^0\right)\,V(r)\,,
\label{eq:harmonicpotential}
\end{equation}
where $V(r)=(ar^2+V_0)$.
In the given expression, $r$ represents the relative distance between quark and antiquark, $\gamma^0$ denotes the time-like Hermitian matrix, and the parameters $a$ and $V_0$ are associated with the potential parameters~\footnote{These parameters have been determined in prior applications of the model through hadron spectroscopy~\cite{Patnaik:2019jho,Patnaik:2023efe}.}.

The internal dynamics of the constituent quarks are assumed to be described by a quark Lagrangian density
\begin{equation}
{\cal L}^{0}_{q}(x)={\bar \psi}_{q}(x)\;\left[\frac{i}{2}\,\gamma^{\mu}
\partial _\mu - m_q-U(r)\;\right]\;\psi _{q}(x)\,,
\end{equation}
leading to the Dirac equation for individual quark
\begin{equation}
\left[\gamma^0E_q-{\vec \gamma}.{\vec p}-m_q-U(r)\right]\psi_q(\vec r) = 0\,,
\end{equation}
with $\psi_q(\vec r)$ representing the four-component Dirac normalized wave function~\cite{Patnaik:2023efe}.

The RIQM is a QCD-inspired phenomenological model designed to characterize the confinement of constituent quarks within hadrons through an interaction potential with a specified Lorentz structure. This model aims to derive observable properties of composite hadrons from the dynamics of constituent quarks, as direct derivation from first principles of QCD is currently unattainable due to inherent complexities. The chosen potential, represented in \eqref{eq:harmonicpotential}, is assumed to depict non-perturbative multi-gluon interactions. Additionally, residual interactions, such as quark-pion coupling from chiral symmetry restoration in the SU(2) sector and one-gluon exchange at short distances, are treated perturbatively in this model.

The selected confining potential yields a simple and manageable form, facilitating the analysis of various hadronic properties and providing an adequate tree-level description for decays involving radiative, rare radiative, leptonic, non-leptonic, and semileptonic modes. The model's applicability and reliability have been validated across a broad spectrum of hadronic phenomena, including radiative, weak radiative, rare radiative~\cite{Barik:1992pq,Barik:1994vd,Priyadarsini:2016tiu,Barik:1995sq,Barik:1996kn,Barik:2001gr}, leptonic~\cite{Barik:1993yj}, weak leptonic~\cite{Barik:1993aw}, semileptonic~\cite{Barik:1996xf,Barik:1997qq,Barik:2009zz}, radiative leptonic~\cite{Barik:2008zza,Barik:2008zz,Barik:2009zza}, and non-leptonic~\cite{Barik:2001vp,Barik:2009zzb,Naimuddin:2012dy,Kar:2013fna}.
This validation spans both light and heavy flavor sectors~\cite{Patnaik:2017cbl,Patnaik:2018sym,Patnaik:2019jho}, affirming the model's broad applicability and robustness. 

The invariant transition matrix element for $B_c\rightarrow \eta_c(J/\psi)l^-\bar{\nu}_l$ and $B_c\rightarrow D(D^*)l^-\bar{\nu}_l$ is expressed as~\cite{Nayak:2021djn}
\begin{equation}
{\cal M}(p,k,k_l,k_\nu)={\frac{\cal G_F}{\sqrt{2}}}V_{bq^{'}}\,{\cal H}_\mu(p,k) \,{\cal L}^\mu(k_l,k_\nu)\,,
\end{equation}
with ${\cal G}_F$ being the effective Fermi
coupling constant, $V_{bq^{'}}$ being the  relevant CKM parameter, ${\cal L}^\mu$ and ${\cal H}_\mu$ are leptonic and hadronic current, respectively. Here, $p,k,k_l,k_\nu$ denote parent ($B_c$) and daughter ($X$) meson's, lepton and neutrino four-momentum, respectively. 

In the context of decay processes, the physical realization depends on the mesons being in their momentum eigenstates. Therefore, a field-theoretic depiction of any decay process requires representing meson bound states through appropriate momentum wave packets. These packets reflect the momentum and spin distribution between the constituent quark and antiquark within the meson core. In the RIQM approach, the wave packet representing a meson bound state, such as $\vert B_c(\vec{p}, S_{B_c})\rangle$ at a specific momentum $\vec{p}$ and spin $S_{B_c}$, is formulated as follows~\cite{Barik:1986mq}
\begin{equation}
\big\vert B_c \left(\vec{p},S_{B_c}\right)\big\rangle = \hat{\Lambda}\left(\vec{p},S_{B_c}\right)\big\vert \left(\vec{p_b},\lambda_b\right);\left(\vec{p_c},\lambda_c\right)\big\rangle \,.
\end{equation}
In the given expression, $\vert (\vec{p_b},\lambda_b);(\vec{p_c},\lambda_c)\rangle$ denotes the Fock space representation of the unbound quark and antiquark in a color-singlet configuration with their respective momentum and spin, and ${\hat \Lambda}(\vec {p},S_{B_c})$ represents an integral operator encapsulating the bound state characteristics of a meson. We then compare our (RIQ) model results with other theoretical approaches and available experimental data.
\section{General kinematics
for \texorpdfstring{$B_c\rightarrow (\eta_c, J/\psi, D, D^*) l \nu_l$ with $l = e, \mu$, $\tau$}{}, and Helicity amplitudes}
\label{sec:testsofLFUV}
In this section, we introduce the $\mathcal{R}$ ratios for $b \rightarrow \tau, \mu$ leptons and examine the longitudinal polarization of leptons in the context of $B_c \rightarrow X l \nu_l$, where $X = \eta_c , J/\psi$, $D(D^*)$.
The angular convention for the transitions is defined in Figure~\ref{fig:ad} and the corresponding Feynman diagram is depicted in Figure~\ref{fig:1}.
The $J/\psi$ is characterized by its purely electromagnetic decay to $\mu^+ \mu^-$, defining the angle $\theta_{V}$.

The study of $B_c$ mesons is particularly intriguing due to their unique status as the lowest bound state. This results in a comparatively long lifetime and a rich spectrum of weak decay channels with significant branching ratios. Consequently, the $B_c$ meson serves as a unique window into heavy quark dynamics, providing an independent test of QCD. With these considerations, we present a model-dependent discussion to address the substantial discrepancies between SM and BSM physics.

By incorporating the weak form factors derived through the covariant expansion of hadronic amplitudes based on the model dynamics, the angular decay distribution in $q^2$ is obtained. Here, $q = p - k = k_l + k_\nu$, and the decay distribution is calculated within the allowed kinematic range: $0 \leq q^2 \leq (M-m)^2$~\cite{Nayak:2021djn}
\begin{equation}
\frac{d\Gamma}{dq^2 d\cos\theta} = {\frac{{\cal G}_F}{(2\pi)^3}}|V_{bq}|^2\frac{(q^2-m_l^2)^2}{8M^2 q^2}|\vec{k}|{\cal L}^{\mu \sigma}{\cal H}_{\mu \sigma}\,.
\label{eq:10}
\end{equation}
In the above ${\cal L}^{\mu \sigma}$ and ${\cal H}_{\mu \sigma}$ represent the lepton and hadron correlation functions, respectively.  
Utilizing the completeness property, the lepton and hadron tensors in Eq.~\eqref{eq:10} are given as 
\begin{eqnarray}
{\cal L}^{\mu \sigma}{\cal H}_{\mu\sigma}              &=&{\cal L}_{\mu'\sigma'}g^{\mu'\mu}g^{\sigma'\sigma}{\cal H}_{\mu\sigma}\,,\nonumber\\
&=&{{\cal L}_{\mu'\sigma'}}{\epsilon^{\mu^{'}}}(m){\epsilon^{\mu^{\dagger}}}(m^{'}){g_{mm^{'}}}{\epsilon^{\sigma^{\dagger}}}(n){\epsilon^{\sigma^{'}}}(n^{'})g_{nn^{'}}{\cal H}_{\mu\sigma}\,,\nonumber\nonumber\\
&=&{L(m,n)}{g_{mm'}}{g_{nn'}}H({m'}{n'})\,.
\end{eqnarray}
where $g_{mn}$ is metric tensor, $g_{mn}={\rm diag}(+,-,-,-)$. For convenience, the lepton and hadron tensors are introduced in the space of helicity components with appropriate four covariant helicity projections~\cite{Nayak:2021djn}
\begin{eqnarray}
L(m,n) &=& {\epsilon^{\mu}}(m){\epsilon^{\sigma ^\dagger}}(n){\cal L}_{\mu \nu}\,,\nonumber\\
H(m,n) &=&{\epsilon^{\mu^\dagger}}(m){\epsilon^{\sigma}(n)}{\cal H}_{\mu \nu}\,.
\end{eqnarray}
The three helicity projections are $m=\pm,0$ which constitute to spin 1 part. The fourth projection is the spin 0, $m=t$ which is the time-component. The polarization vectors are represented by $\epsilon$.

Accordingly, the helicity form factors are defined using Lorentz invariant form factors representing the decay amplitudes, and their evaluation is performed, based on the model dynamics. The Lorentz contraction in Eq.~\eqref{eq:10} is carried out with the helicity amplitudes.

In this study, we omit the consideration of the azimuthal $\chi$ distribution of the lepton pair, thereby integrating over the azimuthal angle dependence of the lepton tensor. The resulting differential distribution in $(q^2, \cos\theta)$ is then obtained as
\begin{eqnarray}
\frac{d\Gamma}{d{q^2}\cos\theta}&=&\frac{3}{8}(1+\cos^2\theta)\frac{d\Gamma_U}{dq^2}+\frac{3}{4}\sin^2\theta \frac{d\Gamma_L}{dq^2}\nonumber\\
&&\mp \frac{3}{4} \cos\theta \frac{d\Gamma_P}{dq^2}+\frac{3}{4}\sin^2\theta\frac{d{\tilde{\Gamma}_U}}{dq^2}+\frac{3}{2}\cos^2\theta\frac{d{\tilde{\Gamma}_L}}{dq^2}\nonumber\\
&&+\frac{1}{2}\frac{d{\tilde{\Gamma}}_S}{dq^2}+3\cos\theta\frac{d{\tilde{\Gamma}}_{SL}}{dq^2}\,.
\label{eq:dcos}
\end{eqnarray} 
The upper and lower signs in Eq.~\eqref{eq:dcos} correspond to the parity-violating term. Among the seven terms, four are identified as $\tilde{\Gamma}_i$, which have significant contribution to non-vanishing lepton mass limit. While the remaining terms are giving dominant contribution over $\tilde{\Gamma}_i$ in vanishing and non-vanishing lepton mass limit which are denoted as $\Gamma_i$. The relationship between them incorporates a flip factor $m_l^2 / (2q^2)$
\begin{equation}
\frac{d{\tilde{\Gamma}}_i}{dq^2}=\frac{m_l^2}{2q^2}\frac{d{\Gamma_i}}{dq^2}\,,
\end{equation}
where the flip factor is expected to significantly influence $\tau$-modes. Thus, the $\tilde{\Gamma}_i$ terms are critical for evaluating the lepton mass effects in semileptonic decay modes. The differential partial helicity rates $d\Gamma_i / dq^2$ are defined by
\begin{equation}
\frac{d\Gamma_i}{dq^2}=\frac{{\cal G}_f^2}{(2\pi)^3}{\vert V_{bq^{'}}\vert}^2 \frac{({q^2}-{m_l^2})^2}{12{M^2}q^2}|\vec{k}|H_i\,.
\label{eq:dw}
\end{equation}
Here, $H_i$ ($i=U,L,P,S,SL$) represents unpolarized-transversed, longitudinal, parity-odd, scalar, and scalar-longitudinal interference helicity structure functions, respectively. These functions are linear combinations of the helicity components of the hadron tensor, $H(m,n) = H_m H_n^\dagger$.
\begin{eqnarray}
\text{Unpolarized-transversed}&:&{H_U}={\rm Re}\left({H_+}{H_+^\dagger}\right)\nonumber\\
&&\quad \quad +\,{\rm Re}\left({H_-}{H_-^\dagger}\right)\nonumber\\
\text{Longitudinal}&:&{H_L}= {\rm Re}\left({H_0}{H_0^\dagger}\right)\nonumber\\
\text{Parity-odd}&:&{H_P}={\rm Re}\left({H_+}{H_+^\dagger}\right)\nonumber\\
&&\quad \quad -\,{\rm Re}\left({H_-}{H_-^\dagger}\right)\nonumber\\
\text{Scalar}&:&{H_S}=3{\rm Re}\left({H_t}{H_t^\dagger}\right)\nonumber\\
\text{Scalar - Longitudinal\ Interference}&:&{H_{SL}}={\rm Re}\left({H_t}{H_0^\dagger}\right)\nonumber
\end{eqnarray}
For spin 1 final state mesons, the contributions include $H_U$, $H_L$, and $H_P$, while for spin 0 final state mesons, they consist of $H_L$, $H_S$, and $H_{SL}$. Assuming the helicity amplitudes to be real due to the $q^2$ range constraint ($q^2 \leq (M-m)^2$), which is below the physical threshold $q^2 = (M+m)^2$, we disregard angular terms multiplied by coefficients ${\rm Im}(H_i H_j^*)$ for $i \neq j^*$. Integrating over $\cos\theta$, we derive the differential $q^2$ distribution. Finally, integrating over $q^2$, the total decay rate $\Gamma$ is obtained as the sum of the partial decay rates: $\Gamma_i$ ($i=U, L, P$) and $\tilde{\Gamma}_i$ ($i=U, L, S, SL$). 

Then, we delve into examining the LFU ratios of $\eta_c$, $J/\psi$, $D$, and $D^*$ in light of their recent lattice predictions. Alongside these ratios, we also investigate angular observables pertinent to these decay processes, namely the lepton polarization asymmetry and the Forward-Backward asymmetry of the lepton, whose definitions were outlined in Section \ref{sec:physical_obs}. Recently, the polarization measurements of final state tau leptons have been proposed as valuable tools for probing BSM physics across various processes. This includes determining the longitudinal polarization of leptons in $B_c \rightarrow \eta_c (J/\psi) D (D^*)$ decays. In addition to studying the ratios $\mathcal{R}(X)$, the longitudinal polarization ($P_L$) of the leptons, the fraction of longitudinal polarization $\mathcal{R}(P_L)$ and the fraction of forward-backward asymmetry $\mathcal{R}_{\mathcal{A}_{\rm FB}}$
are additional physical observables sensitive to specific types of new physics \cite{Harrison:2020nrv, Hu:2019qcn, Wang:2012lrc}. Therefore, we extend our analysis to include similar observables for $\eta_c$, $J/\psi$, $D$, and $D^*$ within the RIQM framework. The results and predictions of these observables are presented in the following section.
\section{Numerical analysis and results}
\label{sec:num_res}
In this section, we present our numerical analysis for exclusive semileptonic decays of $B_c$ mesons into charm and charmonium sectors. The mechanisms governing these decay processes are comprehensively elucidated within the framework of an appropriate phenomenological model, comparing its predictions on observables with those derived from conventional theoretical approaches.

To commence, we compile the input parameters including relevant quark masses, which are determined from the model dynamics as detailed in Table \ref{table:1}. Additionally, we utilize the RIQ model parameters $(a, V_0)$, and the quark binding energy $E_q$ as specified in \cite{Barik:1986mq,Barik:1993aw} 
\begin{eqnarray}
	(a, V_0)=&&(0.017166\ {\rm GeV}^3,-0.1375\ {\rm GeV})\,,\nonumber\\
	(E_b,E_c,E_u)=&&(4.76633,1.57951,0.47125)\ {\rm GeV}\,.
\end{eqnarray}
The meson masses can be found in Table~\ref{table:1}. The values for $V_{\rm CKM}$ are as follows: $V_{\rm cb} = 0.0408 \pm 0.0014$, $V_{\rm ub} = 0.00382 \pm 0.00020$. The lifetime of the $B_c$ meson, $\tau_{B_c} = 0.51 \pm 0.009 \ \text{ps}$, is sourced from \cite{ParticleDataGroup:2022pth}. 
\begin{table}[hbt!]
 \centering
 \setlength\tabcolsep{2.8pt}
 \renewcommand{\arraystretch}{2.0}
\begin{tabular}{|c|c|c|c|c|c|c|c|}
 \hline
 \hline \multicolumn{2}{|c|}{Leptons}& \multicolumn{2}{|c|}{Quarks}&\multicolumn{4}{|c|}{Mesons}\\
 \hline
$e$&0.511&u&78.75&$B_c$&$6274.47\pm0.32$ &$D$&$1864.84\pm 0.05$\\
 $\mu$&105.65&c&1492&$\eta_c$&$2983.9\pm 0.04$&$D^*$&$2006.85\pm0.05$\\
 $\tau$&1776.86&b&4776 &$J/\psi$&$3096.9\pm 0.006$&&\\
 \hline
 \hline 
 \end{tabular}
 \caption{Mass as input parameters (in MeV)}
 \label{table:1}
 \end{table}
The model parameters and quark masses were initially determined in the RIQ model by fitting them to data from heavily flavored mesons, thereby accurately reproducing hyperfine mass splittings within the heavy flavor sector.

Using these input parameters, the Lorentz invariant form factors $(F_+, F_-, A_0, A_+, A_-, V)$ representing the decay amplitudes can be calculated via the overlap integral of participating meson wave functions, taking into account the confinement of constituent quarks within the hadron core through an interaction potential with an appropriate Lorentz structure. The model expressions for these form factors are provided in Appendix \ref{app}. Additionally, we computed these form factors in our approach without employing any extrapolation techniques, thus rendering relatively notable uncertainties in our predictions. A detailed investigation of their $q^2$-dependence and behavior within the allowed kinematic range $q^2_{\rm min} \leq q^2 \leq q^2_{\rm max}$ is presented in \cite{Nayak:2021djn}. 

In this analysis, our focus is exclusively on the $\mathcal{R}$ ratios and angular observables such as the longitudinal polarization of charged leptons, longitudinal $\tau$ polarization, and forward-backward asymmetry, which serve as complementary tests for new physics in specific $B_c$ semileptonic transitions. Given the current absence of predictions in the $B_c$ sectors and the forthcoming data from Run 3, these angular observable predictions provide a notable perspective on the flavor characteristics of heavy flavored mesons.

We first start analyzing the LFU observable $\mathcal{R}$ within our model framework. 
Our predicted values for $\mathcal{R}(\eta_c)$, $\mathcal{R}(J/\psi)$, $\mathcal{R}(D)$, and $\mathcal{R}(D^*)$ are similar in order of magnitude with other model predictions and therefore, are qualitatively in consistent with other SM and lattice predictions \cite{Cooper:2021bkt,Harrison:2020gvo}, see Tables~\ref{table:2} and~\ref{table:2D}. In our current analysis, we specifically consider uncertainties stemming from meson masses, $V_{\rm CKM}$ parameters, and the lifetime of $B_c$. In our findings, the error margins are determined by considering the highest and lowest uncertainties of the input parameters, respectively and then taken the shift from the central values. We have not included uncertainties from our model framework i.e the quark masses, potential parameters or binding energies used in our calculations. It's crucial to highlight that our model does not employ any adjustable free parameters. The potential parameters ($a$, $V_0$), quark masses ($m_q$), and corresponding binding energies ($E_q$) have been fixed from hadron spectroscopy during the foundational application of this model, primarily to reproduce hyperfine splittings of baryons and mesons. This fixed set of parameters is consistently utilized across a wide range of hadronic phenomena, as previously discussed. Consequently, there are no free parameters available within our model framework to introduce uncertainties. 
Our goal in this study is to test the applicability of RIQ model and do a qualitative analysis by comparing our predictions with other phenomenological predictions and the experimental data.
Therefore, the primary theoretical uncertainties in our predictions stem solely from the uncertainties in the input parameters: $V_{\rm cb} = 0.0408 \pm 0.0014$, $V_{\rm ub} = 0.00382 \pm 0.00020$, and $m_{\tau} = 1776.86 \pm 0.12$ MeV. We have dropped these uncertainties in the $\mathcal{R}$ ratios, as the $V_{CKM}$ drops out and there are few dominant source of errors in these observables.
The central values predictions of the $\mathcal{R}$-ratios for the radially excited states in view of the optimised predictions of lepton universality tests in $\psi (2S)$ states \cite{Isidori:2020eyd} is also given in \cite{Nayak:2022gdo}.

Our predicted value of $\mathcal{R}_{J/\psi} = 0.21$ stands approximately 2$\sigma$ below the measurement by LHCb \cite{LHCb:2017vlu} and approximately 1.4$\sigma$ below that of CMS \cite{CMS24:2024CMS}. Should future precision experiments confirm significantly lower central values, it may suggest a violation of LFU in semileptonic $B_c \to J/\psi$ decays. However, the current precision of experimental data is insufficient to definitively support such a conclusion. A more compelling argument could be constructed by incorporating $\mathcal{R}_{\eta_c}$, $\mathcal{R}_{D}$, and $\mathcal{R}_{D^*}$ in $B_c$ decays. In particular, the higher value of $\mathcal{R}_{D} = 0.81$, as shown in Table~\ref{table:2D}, compared to lattice predictions, reflects the substantial contribution of the spin-no-flip component in longitudinal helicity functions, especially more so than in $\mathcal{R}_{\eta_c}$. 
Without experimental observations of LFU ratios in $\eta_c$, $D$, and $D^*$ final states, our phenomenological predictions could prove valuable insights into analyzing the flavor anomalies and potentially identifying $B_c$ channels in the forthcoming Run 3 data at the LHCb experiment.

\begin{table*}[!hbt]
 \centering
\setlength\tabcolsep{4.5pt} 
\caption{Results of ratios of branching fractions for Semileptonic $B_c \to \eta_c (J/\psi)$ decays in the ground state\\ (RIQM, being an approximate model, does not incorporate model parameters uncertainties.)}
\begin{tabular}{|c|c|c|c|c|c|}
\hline Ratio of Branching fractions ($\mathcal{R}$) ($l = e, \mu$) & RIQM & CQM \cite{Issadykov:2017wlb} & PQCD \cite{Wang:2012lrc} & LQCD \cite{Harrison:2020nrv}&LCSR \cite{Leljak:2019eyw}\\
\hline $\mathcal{R}_{\eta_c}=\frac{{\cal B}(B_c\rightarrow \eta_c \tau \nu_{\tau})}{{\cal B}(B_c\rightarrow \eta_c l \nu_l )}$ & 0.43 &0.26&0.34 & -  &0.32 $\pm$ 0.02  \\
\hline $\mathcal{R}_{J/\psi}=\frac{{\cal B}(B_c\rightarrow J/\psi \tau \nu_{\tau})}{{\cal B}(B_c\rightarrow J/\psi l \nu_l )}$ & 0.21  & 0.24&0.28 & 0.2582(38) & 0.23 $\pm$ 0.01\\
\hline
\end{tabular}
\label{table:2}
\end{table*}

\begin{table*}[!hbt]
\centering
\setlength\tabcolsep{5pt} 
\caption{Results of ratios of branching fractions for Semileptonic $B_c \to D (D^*)$ decays in the ground state \\ (RIQM, being an approximate model, does not incorporate model parameters uncertainties.)}
\begin{tabular}{|c|c|c|c|}
\hline Ratio of Branching fractions ($\mathcal{R}$) & RIQM & CQM \cite{Issadykov:2017wlb} & LQCD\cite{Cooper:2021bkt}\\

\hline$\mathcal{R}_D=\frac{{\cal B}(B_c\rightarrow D \tau \nu_{\tau})}{{\cal B}(B_c\rightarrow D \mu \nu_{\mu})}$ & 0.81  & 0.63 & 0.682(37) \\

\hline$\mathcal{R}_{D^*}=\frac{{\cal B}(B_c\rightarrow D^* \tau \nu_{\tau})}{{\cal B}(B_c\rightarrow D^* \mu \nu_{\mu})}$ & 0.91 & 0.56 &-\\
\hline
\end{tabular}
\label{table:2D}
\end{table*}

\begin{table*}[hbt!]
\centering
\setlength\tabcolsep{1.3pt}
\renewcommand{\arraystretch}{2.0}
\caption{Predictions of Longitudinal $\tau$-polarization ($P_{\tau}$) for $\eta_c$, $J/\psi$, $D$ and $D^*$}
\begin{tabular}{|c|c|c|c|}
 \hline $P_{\tau}$ & RIQM & PQCD~\cite{Hu:2019qcn}  & Lattice + PQCD~\cite{Hu:2019qcn} \\
\hline $P_{\tau}\; (\eta_c)$ &  -0.28 $\pm$ 0.0001 & 0.37  $\pm$ 0.01 & -0.36 $\pm$ 0.01  \\
\hline $P_{\tau}\; (J/\psi)$  & -0.56 $\pm$ 0.0003 & -0.55 $\pm$ 0.01 & -0.53 $\pm$ 0.01 \\
\hline $P_{\tau}\; (D)$ & -0.47 $\pm$ 0.0001 & - & - \\
\hline $P_{\tau}\; (D^*)$  & 0.14 $\pm$ 0.0004  & - & -  \\
\hline
\end{tabular}
\label{table:4tau}
\end{table*}

\begin{figure}
\begin{minipage}{0.4\textwidth}
\includegraphics[width=0.94\textwidth]{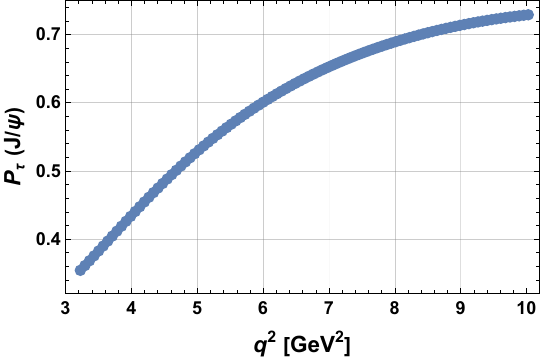}
\includegraphics[width=0.94\textwidth]{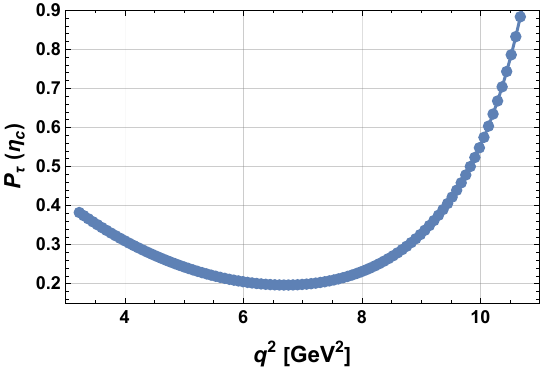}
\includegraphics[width=0.94\textwidth]{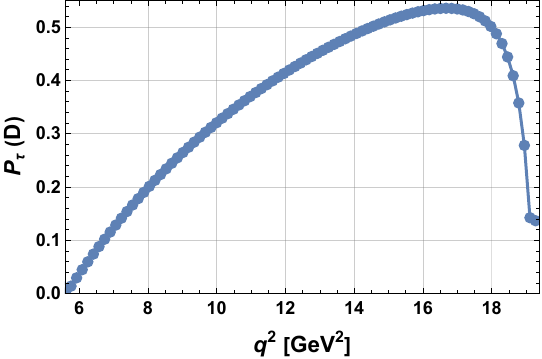}
\includegraphics[width=0.94\textwidth]{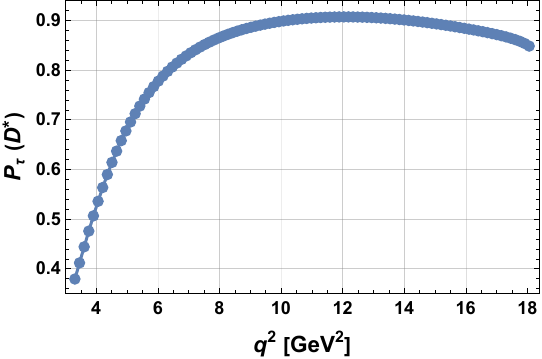}
\end{minipage}
\caption{$\tau$-polarization asymmetry for the $B_c\rightarrow J/\psi$, $B_c\rightarrow \eta_c$, $B_c\rightarrow D$, $B_c\rightarrow D^*$ semileptonic decays in the full $q^2$ kinematic region.}
\label{fig:A}
 \end{figure}

\begin{table}[hbt!]
 \centering
 \setlength\tabcolsep{5pt}
 \renewcommand{\arraystretch}{1.5}
 \caption{Predictions of longitudinal $\mu$ polarization for $B_c$ to charmonium \& charm final states}
\begin{tabular}{|c|c|}
\hline $P_L$ Parameters & RIQM \\
\hline $B_c \rightarrow D (\mu)$, $P^{D}_{\mu}$ & -0.997 $\pm$ 0.00001 \\
\hline $B_c \rightarrow D^* (\mu)$, $P^{D^*}_{\mu}$ & -0.985 $\pm$ 0.00001 \\
\hline $B_c \rightarrow \eta_c (\mu)$, $P^{\eta_c}_{\mu}$ &-0.975 $\pm$ 0.00001 \\
\hline $B_c \rightarrow J/\psi (\mu)$ $P^{J/\psi}_{\mu}$& -0.987 $\pm$ 0.00001 \\
\hline
\end{tabular}
\label{table:5}
\end{table}
Theoretical predictions for the longitudinal polarization of the $\tau$ lepton in the considered semileptonic $B_c \rightarrow (\eta_c, J/\psi) l \nu_l$ decays have been established using two approaches: PQCD and PQCD + Lattice \cite{Hu:2019qcn}. Based on these established theoretical frameworks, we present our results for the longitudinal $\tau$ polarization $(P_{\tau})$ as well as the longitudinal polarization of $\mu$ leptons in Tables \ref{table:4tau} and \ref{table:5} for $\eta_c$ and $J/\psi$ final states. Our predictions for $P_{\tau}(J/\psi)$ are consistent with the predictions derived from LQCD \cite{Harrison:2020nrv,Harrison:2020gvo}, see Figure~\ref{fig:A}. Moreover our prediction of $P_{\tau}(\eta_c) = -0.28$ is also qualitatively similar with the PQCD+Lattice approach \cite{Hu:2019qcn}. Additionally, we provide the results of $\tau$-polarization for semileptonic $B_c\rightarrow D, D^*$ decays. Given the absence of lattice predictions for $\tau$-polarization in $B_c\rightarrow D, D^*$ decays, our model predictions could be instrumental in understanding the angular observables in charm states of $B_c$ semileptonic decays.

It should be noted that for pseudoscalar mesons ($\eta_c$ and $D$), the unpolarized-transverse helicity function does not contribute. The primary contribution for $\eta_c$ and $D$ comes from the flip component of the scalar current function ($\tilde S$), flip component of longitudinal helicity functions ($\tilde L$) and spin-no-flip component of longitudinal helicity functions ($L$), respectively. Conversely, for vector mesons ($J/\psi$ and $D^*$), the tilde and untilde quantities of unpolarized-transverse ($U$), longitudinal ($L$), and scalar ($S$) helicity functions are involved. The central values of the respective partial helicity rates have been reported in \cite{Nayak:2021djn}.

Our predictions for semileptonic modes involving electrons show insignificant differences compared to muon modes, primarily due to the minimal phase space difference between electron and muon channels. Therefore, we present only the longitudinal polarization of muons in Table \ref{table:5}.

To mitigate uncertainties, it was proposed in \cite{Penalva:2020xup} to focus on the ratios $\mathcal{R}_{P_L}$ and $\mathcal{R}_{\mathcal{A}_{\rm FB}}$, which are defined as
\begin{center}
$\mathcal{R}_{P_L} = \frac{P_{L(\tau)}}{P_{L(\mu,e)}}$\,, \quad $\mathcal{R}_{\mathcal{A}_{\rm FB}} = \frac{\mathcal{A}_{\rm FB(\tau)}}{\mathcal{A}_{\rm FB(\mu,e)}}$\,.
\end{center}
Therefore, we present our predictions in Tables \ref{table:7} and \ref{table:8} for the longitudinal $\tau$ polarization fractions in $B_c \rightarrow J/\psi$, $D^*$, $\eta_c$, and $D$ final states, along with the forward-backward asymmetry fractions ($\mathcal{R}_{\mathcal{A}_{\rm FB}}$) for $B_c \rightarrow J/\psi$ and $D^*$, respectively, with their corresponding uncertainties. Measurements and analyses of these quantities provide more comprehensive information than branching ratios, thus defining complementary tests of LFU.

We then plot the $q^2$ spectra for polarized $\tau$ leptons in $B_c \rightarrow \eta_c$, $J/\psi$, $D$, and $D^*$ channels in Figure~\ref{fig:A}. We observe the conservation of angular momentum in the decay process $B_c \rightarrow \eta_c, D$. Given that both the initial and final hadrons have zero spin, the exchanged virtual particle ($W$ boson) must also have zero helicity. In the center-of-mass (CM) system, this implies a zero spin projection along the axis defined by its three-momentum, which is the same as the axis of the final hadron's three-momentum in the CM system. Consequently, in the CM system, the angular momentum of the final lepton pair, measured along this axis, must be zero. This further implies that the helicity of a final $\tau$ lepton emitted along this direction, corresponding to either $\theta_l = 0$ or $\theta_l = \pi$, must match that of the $\nu_{\tau}$, which is always positive.

For the decay transitions $B_c \rightarrow \eta_c$, the longitudinal $\tau$ polarization spectra begin around $q^2 \approx 2 \,{\rm GeV}^2$ and decrease within the range $2 \le q^2 \le 8\, {\rm GeV}^2$ due to phase space constraints, in contrast to $B_c \rightarrow D$, where it steadily increases with increasing $q^2$. Subsequently, there is a consistent rise in the parameter $P_{\tau}(\eta_c)$ as $q^2$ increases. This observed trend can be attributed to the absence of a parity-odd helicity structure function $(H_p)$ in the pseudoscalar amplitude, which significantly influences the spectral continuity and demonstrates strong $q^2$ dependence as the transition progresses from low to high $q^2$.

For $B_c \rightarrow (J/\psi, D^*)$ transitions, in the high $q^2$ region, the parameter $P_{\tau}$ exhibits a steep increase as $q^2$ rises. This behavior arises from the substantial contribution of the no-flip component of the unpolarized-transverse helicity function. The qualitative shape of the $q^2$ dependence of $P_{\tau}(J/\psi)$ agree with the shape of lattice results as reported in \cite{Harrison:2020nrv}.

 \begin{figure}
\begin{minipage}{0.4\textwidth}
\includegraphics[width=0.94\textwidth]{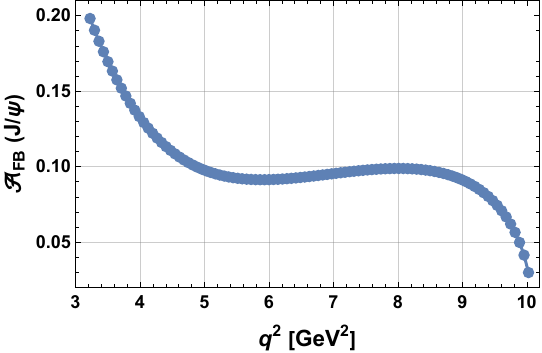}
\includegraphics[width=0.94\textwidth]{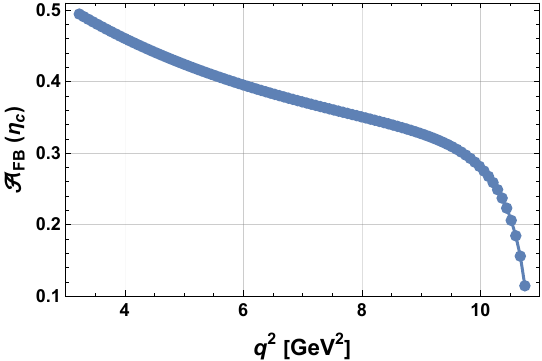}
\includegraphics[width=0.94\textwidth]{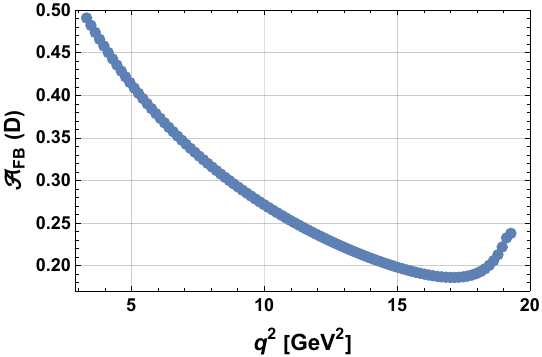}
\includegraphics[width=0.94\textwidth]{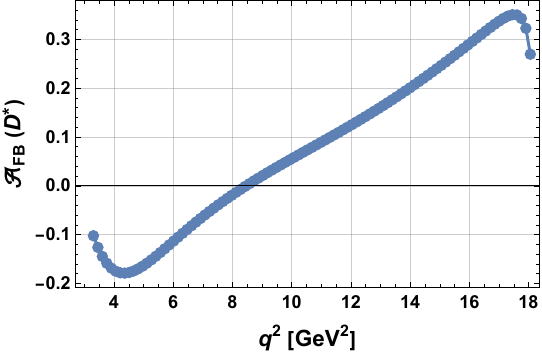}
\end{minipage}
\caption{Forward-Backward asymmetry for the $B_c\rightarrow J/\psi$, $B_c\rightarrow \eta_c$, $B_c\rightarrow D$, $B_c\rightarrow D^*$ semileptonic decays in the full $q^2$ kinematic region.}	
\label{fig:Jpsi}
\end{figure}

The $q^2$ distributions of the forward-backward asymmetry for the polarized $\tau$ lepton in $B_c \rightarrow (J/\psi, \eta_c, D, D^*)$ decays are depicted in Figure~\ref{fig:Jpsi}. For $\eta_c$ and $D$, there is a consistent reduction around the zero recoil region. In the case of $J/\psi$, the $q^2$ distribution decreases initially in the low $q^2$ region until around 6 ${\rm GeV}^2$, followed by a disproportionate reduction in the high $q^2$ region. This reduction is attributed to the longitudinal no-flip amplitude $L$, which decreases due to the presence of a threshold-like factor ${(q^2-m_{\tau}^2)}/{q^2}$ in the rescaled helicity amplitude. This factor affects the amplitude significantly in the high $q^2$ region. However, the shape of the $q^2$ spectra for $\mathcal{A}_{\rm FB}(J/\psi)$ qualitatively agrees with the lattice predictions \cite{Harrison:2020nrv}. When examining the $q^2$ spectra of $\mathcal{A}_{\rm FB}$ for the $B_c \rightarrow D^*$ modes, it is evident that the spin-flip component is minimal compared to the no-flip parts. In this case, the scalar flip component $\tilde{S}$ of the helicity amplitude makes a significant contribution in the high $q^2$ region, thereby influencing the behavior of ${\cal{A}}_{\rm FB}(D^*)$.

\begin{table}[h]
\centering
\setlength\tabcolsep{5pt}
\renewcommand{\arraystretch}{2.0}
\caption{Predictions of longitudinal polarization fraction ($\mathcal{R}_{P_L}$) for $J/\psi$, $D^*$, $\eta_c$ and $D$ final states. (RIQM, being an approximate model, does not incorporate model parameters uncertainties.)}
\begin{tabular}{|c|c|}
\hline $\mathcal{R}_{P_L}$ & RIQM \\
\hline $\mathcal{R}_{P_L} (J/\psi)$  & 0.57 \\
\hline $\mathcal{R}_{P_L} (D^*)$  & 0.138 \\
\hline
$\mathcal{R}_{P_L} (\eta_c)$  & 0.294 \\
\hline$\mathcal{R}_{P_L} (D)$  & 0.468 \\
\hline
\end{tabular}
\label{table:7}
\end{table}

\begin{table}[hbt!]
 \centering
 \setlength\tabcolsep{5pt}
 \renewcommand{\arraystretch}{2.0}
 \caption{Predictions of forward-backward asymmetry fraction ($\mathcal{R}_{\mathcal{A}_{\rm FB}}$) for $J/\psi$, and $D^*$ final states. (RIQM, being an approximate model, does not incorporate model parameters uncertainties.)}
\begin{tabular}{|c|c|}
 \hline $\mathcal{R}_{\mathcal{A}_{\rm FB}}$ & RIQM \\
\hline $\mathcal{R}_{\mathcal{A}_{\rm FB}} (J/\psi)$  & 0.534 \\
\hline
$\mathcal{R}_{\mathcal{A}_{\rm FB}} (D^*)$  & 0.349  \\
\hline
\end{tabular}
\label{table:8}
\end{table}

We give our predictions of $\mathcal{R}_{\mathcal{A}_{\rm FB}}$ for $J/\psi$ and $D^*$ in Table \ref{table:8}. In decays into spin 0 states, $\mathcal{A}_{\rm FB}$ is proportional to the helicity amplitude $\tilde{SL}$, which is negligible in the $e$-mode but non-negligible in transitions involving $\tau$-leptons. For decays into spin-1 states, the transverse component predominates over the longitudinal component by a factor of $\sim$ 2 in the $\tau$-mode for this transition. However, in the $B_c \to D^*$ transition, the transverse component of the partial helicity rates dominates over the longitudinal part by factors of $\sim$ 3 and $\sim$ 7 in the $e$- and $\tau$-modes, respectively.
\section{Summary and outlook}
\label{sec:conclusions}
Exploring the fundamental properties and interactions of particles such as $B_c$ hadrons and $\tau$ leptons has been pivotal in advancing our understanding of elementary particles. This paper investigates a range of angular observables alongside the LFU observables, in $B_c$ mesons, providing insights from both phenomenological and experimental viewpoints. In this analysis, we focused on the $q^2$ distribution of angular observables such as the longitudinal polarization of the final lepton, longitudinal $\tau$ polarization, and forward-backward asymmetry, which serve as complementary tests for BSM physics in $B_c$ semileptonic transitions.

Our predicted value for $\mathcal{R}_{J/\psi}$ is 0.21, which deviates approximately 1.4$\sigma$ below the CMS observation \cite{CMS24:2024CMS} and about 2$\sigma$ below that of LHCb \cite{LHCb:2017vlu}. This suggests a potential violation of LFU in semileptonic $B_c \to J/\psi$ decays, pending further precision experiments to refine central values. However, current experimental data lacks the precision needed to conclusively support such conclusions. Additionally, predictions for $\mathcal{R}_{\eta_c}$, $\mathcal{R}_{D}$, and $\mathcal{R}_{D^*}$ with substantial uncertainties contribute significantly to our analysis. Notably, our prediction of $\mathcal{R}_{D}$ at 0.81 exceeds that from lattice calculations due to the substantial contribution of the longitudinal spin-no-flip component, particularly evident compared to $\mathcal{R}_{\eta_c}$.

The $\tau$ polarization in $B_c \rightarrow \left(\eta_c, J/\psi, D, D^*\right) l \nu_l$ decays serves as a highly sensitive probe of NP, offering valuable insights into the SM. Our predictions for $P_{\tau}(\eta_c, J/\psi)$ aligns with established predictions from PQCD and LQCD in $B_c$ decays. We do qualitative analysis and aim to test the applicability of the RIQ model by demonstrating that our predictions align in terms of order of magnitude with other phenomenological predictions and experimental data, leaving quantitative analysis for the future.
Moreover, the qualitative agreement of $q^2$ distribution spectra for $P_{\tau}(J/\psi)$ and $\mathcal{A}_{\rm FB}(J/\psi)$ with LQCD results is noteworthy in our present analysis. Given the absence of predictions for $P_{\tau}(D, D^*)$, our model's projections stand poised to aid in their analysis as data from Run 3 (2022–2025) becomes available. By integrating LFUV and angular observables into this analysis through a parameter-free unification within our model framework, these results aim to bolster confidence in SM predictions and inform discussions on lepton flavor universality in nature.

Looking forward, with LHCb's completion of Upgrade II and Belle II transitioning into super-B factory operations, the future holds promise for a vibrant era of discovery marked by collaboration and competition among experiments like BESII, ATLAS, and CMS. The exploration of new physics models through experimental testing remains crucial, particularly as we anticipate insights from Run 3 and the Future Circular Hadron Collider (FCC-hh) at CERN, detailed further in Ref.~\cite{Bernlochner:2021vlv}.
\begin{acknowledgments}
S.P. duly acknowledge useful comments, discussions and valuable inputs from Prof. Gino Isidori and the kind hospitality with support of the CERN theory division where this work is completed. S.P. and L.N. are grateful for substantial contribution from Profs. N. Barik, P. C. Dash, and S. Kar in the development of the RIQ model framework.
S.P., L.N., P.S., and S.S. acknowledge NISER, Department of Atomic Energy, India for the financial support.
R.S. acknowledges the support from Polish NAWA Bekker program No. BPN/BEK/2021/1/00342.
\end{acknowledgments}
\begin{widetext}
\appendix
\section{Expressions of the form factors for semileptonic \texorpdfstring{$B_c$}{} to pseudo-scalar (vector) states within the RIQ model}
\label{app}
For $0^-\rightarrow 0^-$ 
transitions, the contribution from the axial vector current is zero whereas for the non-zero vector current components, the spin matrix elements read
\begin{eqnarray}
	\langle S_X(\vec{k})\vert V_0\vert S_{B_c}(0)\rangle =\frac{\left(E_{p_b}+m_b\right)\left(E_{p_{c/u}}+m_{c/u}\right)+|\vec{p_b}|^2}{\sqrt{\left(E_{p_b}+m_b\right)\left(E_{p_{c/u}}+m_{c/u}\right)}}\,,\quad
	\langle S_X(\vec{k})\vert V_i\vert S_{B_c}(0)\rangle =\frac{\left(E_{p_b}+m_b\right)k_i}{\sqrt{\left(E_{p_b}+m_b\right)\left(E_{p_b+k}+m_{c/u}\right)}}\,.
\end{eqnarray}
where $E_{p_b}$ and $E_{p_b+k}$ stand for the energy of the non-spectator quark of the parent and daughter meson, respectively. $E_{p_{c/u}}$ represents the energy of the spectator quark in the $B_c$ transitions to charmonium and charm states. Here $|\vec{p_b}|$ is the respective momentum of the $b$ quark.

Using the above spin matrix elements, the expressions for hadronic amplitudes obtained within the RIQ framework \cite{Nayak:2021djn} are compared with the corresponding covariant expansion expressions, which are formulated in terms of Lorentz invariant form factors. Consequently, the form factors for a pseudo-scalar meson in the final state ($0^- \rightarrow 0^-$) are $f_+$ and $f_-$, which are expressed as follows
 \begin{eqnarray}
 f_\pm (q^2)=\frac{1}{2}\sqrt{\frac{E_k}{M N_{B_c}(0)N_X(\vec{k})}}\int d\vec{p_b}{\cal G}_{B_c}(\vec{p_b},-\vec{p_b}){\cal G}_X(\vec{k}+\vec{p_b},-\vec{p_b})\frac{(E_{o_b}+m_b)(E_{p_{c/u}}+m_{c/u})+|\vec{p_b}|^2\pm(E_{p_b}+m_b)(M\mp E_k)}{E_{p_b}E_{p_{c/u}}(E_{p_b}+m_b)(E_{p_{c/u}}+m_{c/u})}\,.\nonumber
 \end{eqnarray}
 $\mathcal{G}_{B_c}(\vec{p_b},-\vec{p_b})$ and $\mathcal{G}_{X}(\vec{k} + \vec{p_b}, -\vec{p_b})$ are the momentum distribution function of the parent and daughter meson, respectively, derived from the RIQ model dynamics. The detailed expression could be seen in \cite{Patnaik:2017cbl,Patnaik:2018sym,Patnaik:2019jho}.

For $0^- \rightarrow 1^-$ transitions, where the final state is a vector meson, the spin matrix elements corresponding to the vector and axial-vector currents are determined separately and are expressed as
\begin{eqnarray}
 \langle S_X \left(\vec{k},\hat{\epsilon^*}\right)\vert V_0\vert S_{B_c}\left(0\right)\rangle &=&0\,,\\
 \langle S_X \left(\vec{k},\hat{\epsilon^*}\right)\vert V_i\vert S_{B_c}\left(0\right)\rangle&=&\frac{i\left(E_{p_b}+m_b\right)\left(\hat{\epsilon}^*\times \vec{k}\right)_i}{\sqrt{\left(E_{p_b}+m_b\right)\left(E_{p_b+k}+m_{c/u}\right)}}\,,\\
 \langle S_X \left(\vec{k},\hat{\epsilon^*}\right)\vert A_i\vert S_{B_c}\left(0\right)\rangle&=&\frac{\left(E_{p_b}+m_b\right)\left(E_{p_b+k}+m_{c/u}\right)-\frac{|\vec{p_b}|^2}{3}}{\sqrt{\left(E_{p_b}+m_b\right)\left(E_{p_b+k}+m_{c/u}\right)}}\,,\\
 \langle S_X \left(\vec{k},\hat{\epsilon^*}\right)\vert A_0\vert S_{B_c}\left(0\right)\rangle&=&-\frac{\left(E_{p_b}+m_b\right)\left(\hat{\epsilon}^*. \vec{k}\right)}{\sqrt{\left(E_{p_b}+m_b\right)\left(E_{p_b+k}+m_{c/u}\right)}}\,.
 \end{eqnarray}
Furthermore, the expressions for form factors, $V(q^2),A_0(q^2),A_+(q^2)$ and $A_-(q^2)$, within RIQ model read
\begin{eqnarray}
 V\left(q^2\right)&=&\frac{M+m}{2M}\sqrt{\frac{ME_k}{N_{B_c}(0)N_X(\vec{k})}}\int d\vec{p_b}{\cal G}_{B_c}(\vec{p_b},-\vec{p_b}){\cal G}_X(\vec{k}+\vec{p_b},-\vec{p_b}) \sqrt{\frac{\left(E_{p_b}+m_b\right)}{E_{p_b}E_{p_{c/u}} \left(E_{p_{c/u}}+m_{c/u}\right)}}	\,,\\
 A_0(q^2)&=&\frac{1}{(M-m)}\sqrt{\frac{Mm}{N_{B_c}(0)N_X(\vec{k})}}\int d\vec{p_b}{\cal G}_{B_c}(\vec{p_b},-\vec{p_b}){\cal G}_X(\vec{k}+\vec{p_b},-\vec{p_b}) \frac{\left(E_{p_b}+m_b\right)\left(E^0_{p_{c/u}}+m_{c/u}\right)-\frac{|\vec{p_b}|^2}{3}}{\sqrt{E_{p_b}E_{p_{c/u}} \left(E_{p_b}+m_b\right)\left(E_{p_{c/u}}+m_{c/u}\right)}}\,,\\
 A_\pm\left(q^2\right)&=&-\frac{E_k(M+m)}{2M(M+2E_k)}\left[T\mp \frac{3(M\mp E_k)}{(E_k^2-m^2)}\left\{I-A_0(M-m)\right\}\right]\,,
  \end{eqnarray}
  where 
\begin{eqnarray}
E^0_{p_{c/u}}&=&\sqrt{|\vec{p}_{c/u}|^2+m^2_{c/u}} \,, \qquad T=J-\left(\frac{M-m}{E_k}\right)A_0\,,\\
J&=&\sqrt{\frac{ME_k}{N_{B_c}(0)N_X(\vec{k})}}\int d\vec{p_b}{\cal G}_{B_c}(\vec{p_b},-\vec{p_b}){\cal G}_X(\vec{k}+\vec{p_b},-\vec{p_b})\sqrt{\frac{(E_{p_b}+m_b)}{E_{p_b}E_{p_{c/u}}(E_{p_{c/u}}+m_{c/u})}}\,,\\
I&=&\sqrt{\frac{ME_k}{N_{B_c}(0)N_X(\vec{k})}}\int d\vec{p_b}{\cal G}_{B_c}(\vec{p_b},-\vec{p_b}){\cal G}_X(\vec{k}+\vec{p_b},-\vec{p_b})\left\{\frac{(E_{p_b}+m_b)(E^0_{p_{c/u}}+m_{c/u})-\frac{|\vec{p}_b|^2}{3}}{\sqrt{E_{p_b}E^0_{p_{c/u}}(E_{p_b}+m_b)(E^0_{p_{c/u}}+m_{c/u})}}\right\}\,.
\end{eqnarray}
Using the relevant form factors obtained in terms of model quantities, the helicity amplitudes and decay rates for $B_c \rightarrow \eta_c (J/\psi) l \bar{\nu}_l$ and $B_c \rightarrow D (D^*) l \bar{\nu}_l$ transitions are evaluated. Our predictions for these processes are detailed in \cite{Nayak:2021djn}.
 \end{widetext}

\bibliography{sn-bibliography}{}
\bibliographystyle{utphys}
\end{document}